# Why Topology Matters in Predicting Human Activities

Ding Ma[1], Itzhak Omer[2], Toshihiro Osaragi[3], Mats Sandberg[1] and Bin Jiang[1*]

[1]Faculty of Engineering and Sustainable Development
University of Gävle, SE-801 76 Gävle, Sweden
Email: ding.ma|mats.sandberg|bin.jiang@hig.se

[2]Department of Geography and Human Environment
Tel-Aviv University, Tel-Aviv, Israel
Email: omery@post.tau.ac.il

[3]Department of Architecture and Building Engineering
Tokyo Institute of Technology, Tokyo, Japan
Email: osaragi.t.aa@m.titech.ac.jp

*(Draft: December 2017, Revision: February, June 2018)*

**Abstract**
Geographic space is better understood through the topological relationship of the underlying streets (note: entire streets rather than street segments), which enables us to see scaling or fractal or living structure of far more less-connected streets than well-connected ones. It is this underlying scaling structure that makes human activities predictable, albeit in the sense of collective rather than individual human moving behavior. This topological analysis has not yet received its deserved attention in the literature, as many researchers continue to rely on segment analysis for predicting human activities. The segment-analysis-based methods are essentially geometric, with a focus on geometric details of locations, lengths, and directions, and are unable to reveal the scaling property, which means they cannot be used for human activities prediction. We conducted a series of case studies using London streets and tweet location data, based on related concepts such as natural streets, and natural street segments (or street segments for short), axial lines, and axial line segments (or line segments for short). We found that natural streets are the best representation in terms of human activities or traffic prediction, followed by axial lines, and that neither street segments nor line segments bear a good correlation between network parameters and tweet locations. These findings point to the fact that the reason why axial lines-based space syntax, or the kind of topological analysis in general, works has little to do with individual human travel behavior or ways that human conceptualize distances or spaces. Instead, it is the underlying scaling hierarchy of streets – numerous least-connected, a very few most-connected, and some in between the least- and most-connected – that makes human activities predictable.

## 1. Introduction
Human activities can be predicted well by the topology of space, or, more precisely, the topological relationship of named or natural streets that are generated from individual street segments with same names or good continuity (Jiang et al. 2008). The topology of streets enables us to see the underlying scaling of far more less-connected streets than well-connected ones, or the scaling hierarchy of numerous least-connected, a very few most-connected, and some in between the least- and most-connected (Jiang and Yin 2014). Thus, the topology of streets differs fundamentally from the concept of topology in geographic information systems (GIS), since the latter is imposed among geometric primitives of points, lines, polygons, and pixels (Longley et al. 2015). The topology of streets differs also fundamentally from the topology in segment analysis, which is defined among adjacent line segments (Hillier and Iida 2005); see more details in Section 3. Thus, the topology of streets is unique, being a notion that is closely related to the scaling structure of far more smalls than larges; see more details in Section 2. It is this scaling or fractal or living structure that makes human activities predictable (Note: the notion of living is taken from Alexander (2002–2005)), albeit in the sense of collective rather than individual human moving behavior. In other words, human activities or traffic flow captured by the kind of topological analysis, or space syntax in particular (Hillier and Hanson 1984), are collective human travel behavior shaped by the underlying street structure (Penn 2003, Penn et al. 1998). The reason why human activities can be predicted by the topological analysis of streets, or why axial lines-based space syntax works, has little to do with individual human factors such as age, gender, or an individual's familiarity with the space, or ways that human conceptualize space or distances. Instead, it is essentially the underlying street structure – or, more precisely, the underlying scaling structure of far more less-connected than well-connected streets – that shapes human activities (e.g., Jiang and Jia 2011, Omer and Jiang 2015). In other words, given a street network, traffic flow with respect to individual streets for human beings is essentially the same as that for monkeys or random walkers. However, this advantage of topological analysis has not yet received deserved attention in the literature. The present paper intends to illustrate that natural streets are the best topological representation for predicting human activities.

This paper was initially motivated by the basic question of why space syntax works or, more precisely, why axial lines-based space syntax works. The same question has been well addressed in the space syntax literature. For example, Jiang and colleagues (Jiang and Jia 2011, Jiang et al. 2008, Omer and Jiang 2015) concluded that traffic flow captured by space syntax is morphological, which means it has little to do with human factors, as mentioned above. Importantly, axial lines are not a good representation in comparison with natural streets at the city scale. On the other hand, Hillier and colleagues (Hillier 1999, and Hillier and Iida 2005) looked for reasons from both morphological (or network effect) and human factors, particularly different ways in which human beings conceptualize distances. We will argue in this paper that the segment analysis is essentially geometry-based rather than topology-oriented, meaning that it cannot be used for predicting human activities. At this point, the reader may be wondering why good correlations were found by segment analysis-related studies, including the London case studies (Hillier and Iida 2005). One main reason could be the small case studies areas, which do not involve a large number of axial lines or line segments. Also, the correlations through segment analysis are not as good as those in the corresponding axial line analysis (Hillier and Iida 2005). Another reason could be the distance-decay effect while calculating the correlations (Gao et al. 2013). The present paper reports several case studies at the city scale using street or segment representations without considering other factors in order to demonstrate that natural streets are the best representation for predicting human activities, followed by axial lines, and that neither street segments nor line segments are good representation for capturing human activities. This is because the first two representations are topological, while the latter two representations are geometric.

The contribution of this paper is three-fold. First, we illustrate with empirical evidence that human activities are mainly shaped by the underlying scaling of far more less-connected streets than well-connected ones, and that topological analysis capable of capturing the underlying scaling can predict human activities well. Second, we demonstrate that segment analysis or conventional GIS representations are essentially geometric and therefore cannot reveal the scaling or fractal or living structure, which means they are unable to predict human activities. Third, we further clarify the importance of topological analysis in space syntax.

The remainder of this paper is structured as follows. Section 2 introduces some basic principles of topological analysis and space syntax, concentrating on two basic notions of topology and scaling. Section 3 illustrates segment analysis and makes it clear why the approach is geometric in essence. Section 4 reports a series of case studies using London street network and tweet location data, in order to demonstrate that natural streets are the best representation for traffic flow prediction, followed by axial lines, and that neither street nor line segment is able to capture human activities. Section 5 further discusses the study and its implications for better understanding urban structure and dynamics. Finally, Section 6 concludes and points to future work.

## 2. Basics of topological analysis and space syntax

The topological analysis was initially developed from space syntax, which is a set of techniques for urban morphological analysis based on graph theory (Hillier and Hanson 1984, Penn 2003, Penn et al. 1998, Batty 2004). The fundamental principle of space syntax is to represent an urban space topologically rather than geometrically (Figure 1, which was modified from Jiang and Claramunt 2004) using graph theory, or more correctly, complex networks (Watts and Strogatz 1998, Barabási and Albert 1999, Newman 2010, Porta et al. 2006, Blumenfeld-Lieberthal 2009, Saberi et al. 2017). The most commonly used representation is the axial map, consisting of a least number of longest visibility lines (so-called axial lines) covering an entire urban space – the street space or the space between buildings where people can move around freely. The least number of longest axial lines is then converted into a connectivity graph, in which vertices represent axial lines, while edges represent intersections of the axial lines. Based on the binary connectivity graph, a set of morphological parameters – various centrality measures (Harary 1969, Freeman 1979, Hillier and Hanson 1984) – are used to characterize the structure of axial map or of the urban space. It is the connectivity graph or the topology that enables us to see the underlying scaling of far more less-connected streets than well-connected ones. This is one of the major points that we want to make in this paper. The key of space syntax is not just a graph representation, but the one that is able to capture underlying scaling of far more less-connected things than well-connected ones. In this section, we use the concept of natural streets to illustrate the fundamental principles of topological analysis or of space syntax in particular.

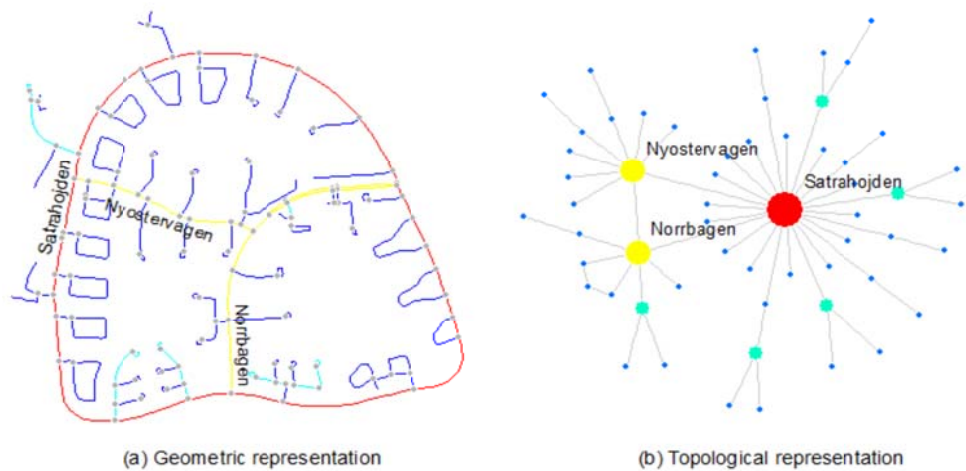

Figure 1: (Color online) Transformation of geometric representation to topological representation, which enables us to see the underlying scaling of far more less-connected than well-connected streets

(Note: Geometric representation (a), due to geometric details of locations, lengths, and directions, must be transformed into the topological representation (b) in order to see clearly the scaling property of far more less-connected streets than well-connected ones. The topological representation bears no geometric details at all. The most distinguished feature of the topological representation is its underlying scaling of numerous least-connected, a very few most-connected (only one), and some in between the least- and most-connected.)

Figure 1a illustrates a neighborhood called Sätra in the Swedish city of Gävle, and it appears to be a street space without buildings or other geographic features represented. It is usually called a "topological model" in GIS or, more specifically, a topologically integrated geographic encoding and referencing (TIGER) model, which was developed by the US Census Bureau in the 1970s. However, this "topological model" is essentially geometric, because of geometric details of locations, lengths and directions. Structurally, it is simple or even boring since every junction or segment has a more or less similar degree of connections. The geometric model is a good model for calculating distances in route planning and many applications (Osaragi 2013, Shen and Karimi 2017), but it is unable to reveal the underlying scaling structure. Transformed directly from the geometric one, Figure 1b is a truly topological model, in which vertices represent individual streets, while the street-street intersections become edges of the graph. It is obvious that the metric information of streets such as lengths and widths is discarded in the transformation. The ring road has the highest degree of connectivity, with 24 other streets directly connected to it. The ring street is the most central in terms of not only degree centrality, but also closeness – close to all other streets. The three longest streets have also the highest degree of betweenness, since they tend to bridge different parts of the connectivity graph. The most distinguished feature of the topological representation is its scaling hierarchy of many least-connected (49), only one most-connected, and some in between the least- and most-connected. There are four hierarchical levels, indicted by the four colors in Figure 1.

Topology and scaling are the two fundamental and interconnected concepts behind the topological analysis or space syntax in particular. The topology helps show the underlying scaling of street connectivity: far more the less-connected than the well-connected. In other words, a street space or urban environment is usually too big or too complex to perceive directly. However, it can be well perceived by partitioning the space into many perceivable pieces, and then connecting these pieces to become a coherent whole: the connectivity graph. These perceivable pieces are not evenly or equally distributed like pixels in an image, or like points, lines, and polygons in the TIGER model. Instead, there are far more small pieces than large ones, or far more short streets than long ones, or far more less-connected things than well-connected ones. Importantly, every piece is small or simple enough to be perceived from a single vantage point. Essentially, a large, complex space can be understood from the connectivity graph, which captures the scaling of far more small pieces than large ones. Unlike topology and scaling, geometric representations concentrate on geometric details of distances, locations, and directions, implying that they are incapable of revealing the underling scaling of far more less-connected things than well-connected ones. Instead, they show more or less similar segments or junctions. In the next section, we introduce one of the geometric representations: so-called segment analysis. The segments initially refer to line segments, but we introduce the segment analysis using street segments in order to consistent with natural streets.

## 3. Geometric representations and segment analysis

Geographic representations of current GIS, such as vector and raster, are essentially geometric rather than topological. We have categorized them as geometric, not because there are no topological relationships, but because the topological relationships recur among geometric primitives of points, lines, polygons, and pixels (e.g., Longley et al. 2015). In the GIS literature, topology or topological relationship is weighted by distances or constrained by geographic locations. The London underground map is a good example of the topology featured in current GIS. Despite the distorted geometric aspects of the map, its topological relationships are defined among individual stations. As a reminder, the topology we refer to enables us to see the scaling of far more small things than large ones (see Section 2 for an illustration). Thus, all geographic representations in conventional GIS are essentially geometric; one can refer to related textbooks for more details (e.g., Longley et al. 2015). In this section, we introduce another geometric representation: so-called segment analysis.

The segment analysis is based on a weighted graph among individual line segments (Hillier and Iida 2005). The authors used axial line segments to build up the graph, but we introduce the method using street segments in order to be consistent with the street-based topological representation (Figure 2). As shown in Figure 2, the three most connected streets consist of six street segments: $s_1, s_2, \ldots, s_6$. Their adjacent relationships constitute a matrix, in which the elements or weights are defined by three different distances: metric, fewest-turn, and angular. The metric distance is defined by the sum of the lengths of the adjacent segments; actually, half the sum of their length between the two mid-points. The fewest-turn distance of two adjacent segments is defined as follows: the distance is one if the two adjacent segments are not at the same direction; otherwise the distance is 0. The angular distance is defined by the sum of angular changes, so it is also commonly called geometric. These three different distances or weights reflect how human beings conceptualize distances, as Hillier and Iida (2005) attempted to seek why space syntax works from human aspects. This focus on human aspects differs fundamentally from what we seek from the underlying street structure or the scaling of far more less-connected things than well-connected ones.

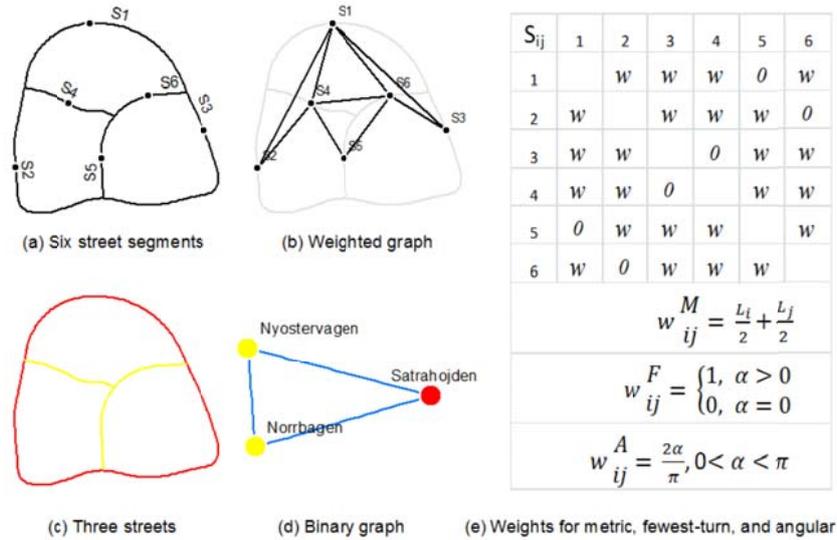

Figure 2: (Color online) Segment analysis versus topological analysis
(Note: Six street segments (a) and their weighted graph (b), which is represented as a weighted matrix (e), where the weights *w* could be metric, fewest-turn, and angular. The six street segments constitute the three natural streets (c), and their topology or binary graph (d), both of which are a simplified version of Figure 1.)

Unlike the topological analysis, segment analysis cannot be used to reveal the underlying scaling of street space. This can be seen from two aspects. First, the concept of segment is mechanistically defined and holds little meaning in our perception or cognition, very much like geometric primitives in conventional GIS representations. Second, all street segments have a more or less similar degree of connectivity, rather than being far more the less-connected than the well-connected. This is the same for street junctions. What makes space syntax unique, or what makes space syntax different from GIS, is the topological perspective that enables us to see the true scaling structure. Therefore, the segment analysis is essentially geometric and differs fundamentally from the topological analysis.

There is an alternative way of establishing segment-segment relationships through natural streets, i.e.: segments $i$ and $j$ are linked if they are on the same natural street. The segment-segment relationships are not just among adjacent segments, but through their individual streets. Unlike the segment analysis above, this segment-based representation is binary. This method is also called "point-based space syntax" (Jiang and Claramunt 2002), which relies on the characteristic points (that is, road junctions and turning points) on the street network to build up a connectivity or visibility graph. These characteristic points can be replaced by segments for constructing the connectivity graph (Jiang et al. 2008). A series of graph parameters such as centrality can be calculated for the segments or points, and the parameters are assigned to individual streets by summation. The aggregated parameters on natural streets possess a striking scaling property and they hold good correlations with human activities. The following experiments will provide related empirical evidence.

## 4. Experiments on predicting human activities
We conducted a series of experiments on correlation analysis between the underlying scaling property of streets and number of tweet locations in order to demonstrate that (1) natural streets are better than axial lines in predicting human activities, and (2) neither street or line segment is able to capture the scaling pattern of human activities, but that segment-based parameters that are aggregated into natural streets or axial lines-based graph can again have good correlations with human activities. These experiments prove that topological rather than geometric representation or the underling scaling structure shapes human activities. Below, we will show that if a study area is small enough, we will be able to find good correlation for segment-based parameters and human activities. This good correlation just occurred by chance, rather than being systematic or universal.

### 4.1 Data and data processing
Relying on the OpenStreetMap data source in London, the natural streets were auto-generated based on the self-best principle (Jiang et al. 2008), and axial lines were also auto-generated based on the notion of drivability (Liu and Jiang 2012). The London city boundary is that of the natural city derived from all Tweet locations of entire UK (Jiang 2017). The generation of natural streets and axial lines can be done by Axwoman (http://giscience.hig.se/binjiang/). Subsequently, we derived the segments from natural streets and axial lines, respectively, by chopping at the street junctions where at least three polylines intersect. Geo-tagged tweet locations were used to indicate human activities, and 165,963 tweet locations were collected through the Twitter Streaming Application Program Interface. The time duration for tweet data is June 1–8, 2014 in London. These tweet locations were then snapped into individual streets, axial lines, and street and line segments respectively for correlations. The snapping was done in ArcGIS software (ArcToolBox > Analysis > Near) by calculating the one-to-one spatially closest relationship between tweet location and each type of street/segment. We used 10 meters as a search tolerance for each tweet location. For each geo-tagged tweet, if there was no street/segment within its 10-meter buffer, the tweet was not used for further analysis. After clipping out the raw street network data, we first ensured the topological correctness of the data, by removing duplicated lines and creating the street junction nodes at the intersection.

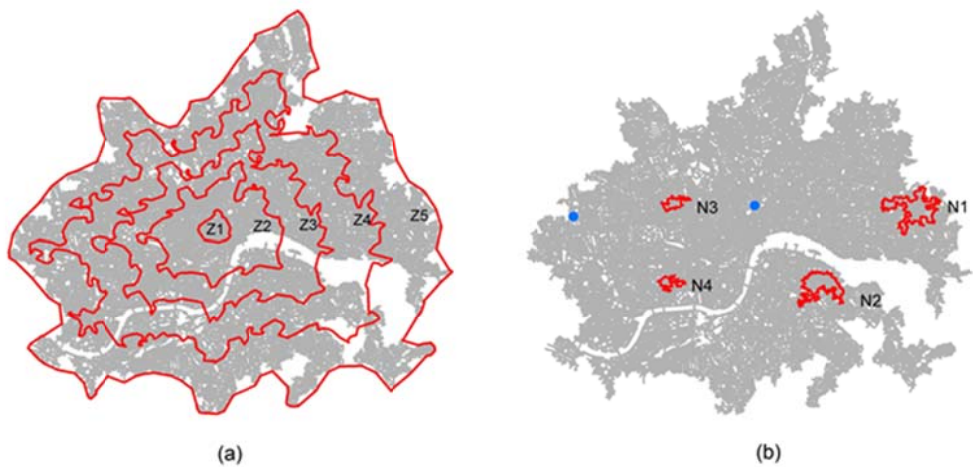

Figure 3: (Color online) The nine study areas of London used for the experiments

There are two types of study area: Five areas (Z1, Z2, Z3, Z4, and Z5) from the city center to the periphery and four areas (N1, N2, N3, and N4) scattered around the city (Figure 3). Note that Z1, Z2, Z3, Z4, and Z5 are nested with each other; that is, small zones are included in the large ones: Z1 ⊂ Z2 ⊂ Z3 ⊂ Z4 ⊂ Z5, where Z5 refers to the entire London area. The four scattered areas (N1, N2, N3, and N4) were not chosen randomly, but had high concentrations of tweets. More specifically, we clustered the tweet locations by aggregating edges that are shorter than the average of all edge lengths under a big triangulated irregular network based on all tweet locations (Jiang and Miao 2015). Using this method, we obtained numerous clusters whose sizes exhibit a strikingly scaling property. From these clusters, we chose two largest popular tweeted areas (N1 and N2), and two middle-sized ones (N3 and N4).

### 4.2 Presence and absence of scaling property
The natural streets and axial lines show very striking scaling property of far more the less-connected than the well-connected, or numerous least-connected (blue), a very few most-connected (red), and some in between the least- and most-connected (other colors) (Figure 4a and b). The classification is based on head/tail breaks – a new classification scheme and visualization tool (Jiang 2013, Jiang 2015). The scaling property is further illustrated using cumulative distribution function in Figure 4c and Figure 4d, and power law fitting metrics in Table 1, where the degree of natural streets passes the strict power law detection (Clauset et al. 2009) with a goodness of fit, p-value ≥ 0.8. However, the connectivity of axial lines is less power law, because the power law exponent is outside of the acceptable range (1 ≥ $\alpha$ ≥ 3). Therefore, the natural streets possess a more strikingly scaling pattern than axial lines. In contrast, neither street segment nor line segment shows a power-law relationship, because alpha values are outside of the acceptable range. To our surprise, the connectivity of street segments shows a very high ht-index of 5. Furthermore, both minimum and maximum of connectivity help us better understand the presence and absence of the scaling property. A wide range between the maximum and minimum connectivity is a condition for the presence of the scaling. On the other hand, street and line segments show some very different statistics. For example, most of the street segments have connectivity of 4 and 5, while most of the line segments have connectivity of 3 and 6. It is probable that the street segments have a high ht-index, while line segments have a very low ht-index.

Table 1: Statistics on connectivity of different street/segment representations in the London area (Note: # = street/segment number, Alpha = power law exponent, Xmin = the minimum connectivity being power law, p-value = index for goodness of fit, max = maximum connectivity value, min = minimum connectivity value, ht = ht-index of connectivity (Jiang and Yin 2014))

|  | # | Alpha | Xmin | p-value | max | min | ht |
|---|---|---|---|---|---|---|---|
| Natural street | 29,609 | 2.77 | 6 | 0.8 | 243 | 1 | 9 |
| Axial line | 49,409 | 3.36 | 8 | 0.01 | 143 | 1 | 9 |
| Street segment | 89,495 | 9.43 | 6 | 0 | 11 | 1 | 5 |
| Axial line segment | 250,652 | 2.66 | 3 | 0 | 14 | 1 | 1 |

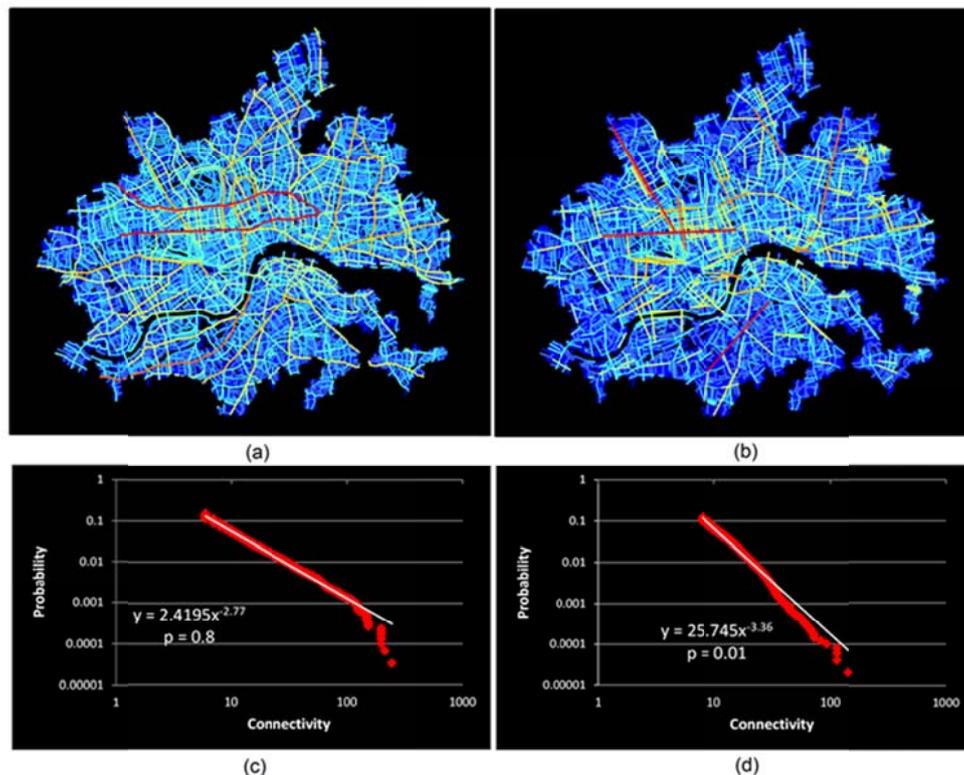

Figure 4: (Color online) The natural streets (a) and axial lines (b) in the London area and the power law distribution plot of the connectivity of natural streets (c) and axial lines (d)
(Note: There are numerous least-connected streets or lines (blue), a very few most-connected (red), and some in between the least- and most-connected (other colors).)

## 4.3 Presence and absence of correlation between connectivity and human activities

Given the scaling property, it is not surprising that there is a good correlation between the connectivity of natural streets and the number of tweet locations. The overall correlation is around 0.4 of R square in the London area (Z5). However, from the outmost border (Z5) to the city center (Z1), the R square increases up to 0.85 (Table 2). This may result from two factors: (1) human activities tend to be higher in the center, and low in the periphery, and (2) there is an edge effect for the topological analysis. To avoid the edge effect, the correlation analyses were examined further in several other regions with a high concentration of tweet locations (N1 to N4). The correlations are also very high (Table 2). By comparison, the correlations for the axial lines in the selected areas are much lower than those for the natural streets. These results indicate that natural streets are better axial lines for predicting human activities.

Table 2: Correlation between the connectivity and the number of Tweet locations in nine study areas
(Note: # = number, NS = Natural street, AL = Axial line, SS = Street segment, LS = Line segment)

| Zone# | Correlation (NS) | Correlation (AL) | Correlation (SS) | Correlation (LS) |
|---|---|---|---|---|
| Z1 | 0.85 | 0.31 | 0.003 | 0.02 |
| Z2 | 0.62 | 0.19 | 0.008 | 0.002 |
| Z3 | 0.57 | 0.14 | 0.008 | 0.001 |
| Z4 | 0.51 | 0.12 | 0.007 | 0.002 |
| Z5 | 0.36 | 0.08 | 0.0002 | 0.001 |
| N1 | 0.39 | 0.02 | 0.002 | 0.0001 |
| N2 | 0.67 | 0.14 | 0.0003 | 0.002 |
| N3 | 0.83 | 0.24 | 0 | 0.005 |
| N4 | 0.89 | 0.53 | 0.01 | 0.007 |

Unlike natural streets and axial lines, street and line segments do not demonstrate the scaling property because they do not pass the power law detection, as the p-values equal to 0 (Table 1). Visually, the street and line segments do not exhibit a clear scaling pattern, and instead show some random structure, as shown in Figure 5. More specifically, the connectivity values of street segments show a normal distribution, whereas those of line segments show a bimodal distribution (Figure 5). The two histograms demonstrate that the connectivity values of street or line segments tend to center around the mean with a small deviation. The absence of scaling property determines the extremely poor correlation between segment connectivity and the number of tweet locations; most R square values for street and line segments are less than 0.01 (Table 2). This is in line with our argument that the underlying scaling structure shapes human activities. So far, we have only examined connectivity, its scaling property, and its correlation with the number of tweet locations. In the next subsection, we will further examine two other centrality measures (betweenness and closeness) with respect to the four segment-based representations.

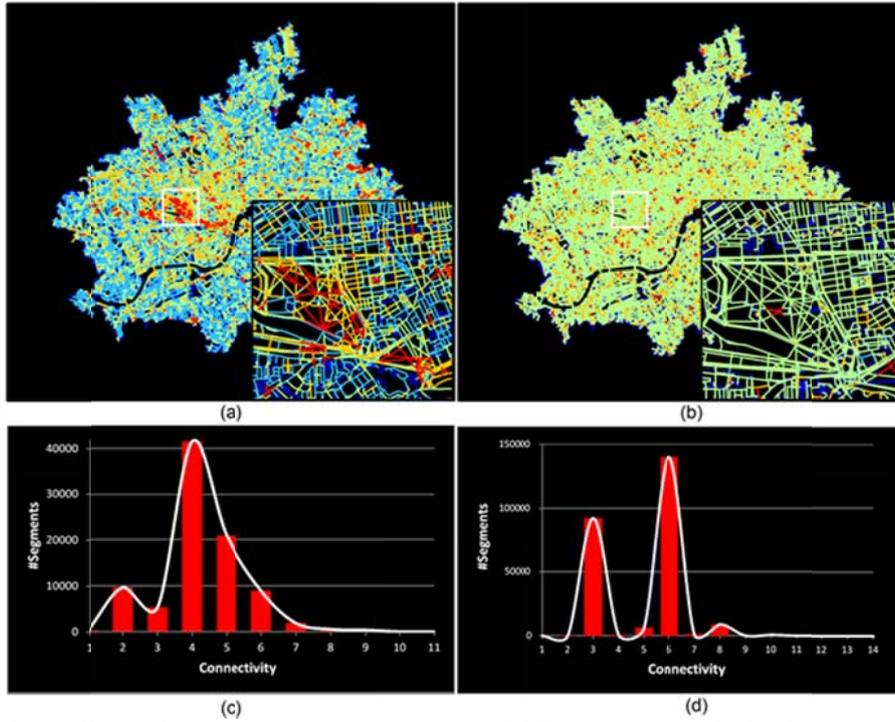

Figure 5: (Color online) The street segments (a) and axial line segments (b) in the London area and the histogram of the connectivity of the street segments (c) and line segments (d)
(Note: The insets in panels a and b provide the enlarged view of street or line segments in central London)

### 4.4 Correlations with other centrality measures based on segment analysis

We further examined correlations to confirm that the segment-based street representations (both street and line segments) fail to capture the underlying scaling pattern of human activities. In order to do this, we computed other two centralities: betweenness and closeness, based on different types of segment-segment graph concerning least length, minimum angle, fewest turns, and natural streets as introduced in Section 3. As shown in Table 3, the entire London and selected areas exhibit very poor correlations (R square < 0.1) among four types of graphs between segment metrics and the number of tweet locations. To be consistent with the previous study (Hillier and Iida 2005), which showed high metric-flow correlations (R square above 0.5) in certain areas, we also found such areas (shown as blue points in Figure 3) using other natural cities' boundaries based on tweet locations where the axial line segment angular-or length-based betweenness correlate very well with the number of tweet locations. But those areas are not representative enough (that is, there is not a high enough concentration of tweets) to be chosen for this study.

Table 3: R square values between the centrality metrics and the number of tweet locations in the study areas under street and line segment analyses
(Note: M = Metric, F = Fewest turns, A = Angular, P = Point-based)

|  |  | Betweenness | | | | Closeness | | | |
|---|---|---|---|---|---|---|---|---|---|
|  |  | M | F | A | P | M | F | A | P |
| Street segment | Z1 | 0.022 | 0.041 | 0.052 | 0.002 | 0.014 | 0.111 | 0.083 | 0.076 |
|  | Z2 | 0.002 | 0.005 | 0.005 | 0.001 | 0.006 | 0.011 | 0.008 | 0.023 |
|  | Z3 | 0.003 | 0.004 | 0.004 | 0.0002 | 0.007 | 0.008 | 0.007 | 0.017 |
|  | Z4 | 0.002 | 0.003 | 0.002 | 0.0001 | 0.004 | 0.004 | 0.004 | 0.01 |
|  | Z5 | 0.002 | 0.002 | 0.002 | 0.0001 | 0.003 | 0.004 | 0.003 | 0.006 |
|  | N1 | 0.003 | 0.001 | 0.0004 | 0.0001 | 0 | 0.001 | 0.001 | 0.002 |
|  | N2 | 0.0002 | 0.001 | 0 | 0 | 0.001 | 0.0002 | 0 | 0.004 |
|  | N3 | 0.001 | 0.016 | 0.035 | 0.001 | 0.006 | 0.009 | 0.015 | 0 |
| | N4 | 0.019 | 0.004 | 0.022 | 0.027 | 0.012 | 0.024 | 0.024 | 0.056 |
| Line segment | Z1 | 0.009 | 0.009 | 0.031 | 0.006 | 0.001 | 0.077 | 0.073 | 0.08 |
| | Z2 | 0.002 | 0.003 | 0.015 | 0.0002 | 0.003 | 0.017 | 0.011 | 0.015 |
| | Z3 | 0.003 | 0.003 | 0.012 | 0.0003 | 0.004 | 0.011 | 0.007 | 0.009 |
| | Z4 | 0.003 | 0.003 | 0.006 | 0.0003 | 0.002 | 0.007 | 0.004 | 0.006 |
| | Z5 | 0.003 | 0.002 | 0.005 | 0.0003 | 0.002 | 0.004 | 0.003 | 0.004 |
| | N1 | 0.001 | 0.001 | 0.002 | 0 | 0.0004 | 0.001 | 0.0001 | 0.001 |
| | N2 | 0.002 | 0.002 | 0.003 | 0.002 | 0.001 | 0.004 | 0.001 | 0.002 |
| | N3 | 0.001 | 0.001 | 0.019 | 0.001 | 0.007 | 0.008 | 0.01 | 0.008 |
| | N4 | 0.019 | 0.013 | 0.002 | 0.037 | 0.006 | 0.014 | 0.014 | 0.024 |

The correlations are significantly improved (R square > 0.6) in some areas (such as Z1 and N4) when we summed the centrality measures of segments into individual natural street; see Table 4. For closeness centrality, the R square values even climb to 0.9. There is also a certain improvement on correlations based on the aggregated values of line segment metrics using axial lines (Table 4). The reason for the high correlation can again be said to be the effect of scaling. Using Z1 as an example, after the segment metrics are summed based on individual natural street or axial line, either betweenness or closeness values among four types of segment graphs have a scaling property (Figure 6). Furthermore, the values from the natural street graph are the most strikingly scaling. In this connection, it is not surprising that the R sqaure values can increase dramatically from less than 0.1 to around 0.9, and the summed values based on natural street graph in most cases show the best correlations. Based on these findings, we further confirm that the human activities in particular are shaped mainly because the underlying scaling structure of urban space.

Table 4: R square values between the summed segment centrality values and the number of tweet locations based on natural streets and axial lines
(Note: M = Metric, F = Fewest turns, A = Angular, P = Point-based)

| | | Betweenness | | | | Closeness | | | |
|---|---|---|---|---|---|---|---|---|---|
| | | M | F | A | P | M | F | A | P |
| Natural streets | Z1 | 0.61 | 0.66 | 0.64 | 0.65 | 0.85 | 0.86 | 0.86 | 0.87 |
| | Z2 | 0.49 | 0.2 | 0.23 | 0.53 | 0.68 | 0.68 | 0.68 | 0.71 |
| | Z3 | 0.44 | 0.1 | 0.16 | 0.34 | 0.64 | 0.64 | 0.64 | 0.67 |
| | Z4 | 0.39 | 0.1 | 0.15 | 0.31 | 0.6 | 0.59 | 0.59 | 0.62 |
| | Z5 | 0.37 | 0.09 | 0.14 | 0.29 | 0.55 | 0.55 | 0.54 | 0.57 |
| | N1 | 0.56 | 0.06 | 0.08 | 0.23 | 0.43 | 0.41 | 0.42 | 0.42 |
| | N2 | 0.48 | 0.29 | 0.42 | 0.52 | 0.68 | 0.7 | 0.7 | 0.72 |
| | N3 | 0.75 | 0.65 | 0.65 | 0.76 | 0.83 | 0.83 | 0.83 | 0.84 |
| | N4 | 0.94 | 0.42 | 0.28 | 0.84 | 0.91 | 0.92 | 0.92 | 0.94 |
| Axial lines | Z1 | 0.37 | 0.07 | 0.28 | 0.46 | 0.42 | 0.41 | 0.42 | 0.41 |
| | Z2 | 0.16 | 0.06 | 0.2 | 0.16 | 0.24 | 0.27 | 0.25 | 0.27 |
| | Z3 | 0.15 | 0.06 | 0.16 | 0.13 | 0.18 | 0.21 | 0.2 | 0.21 |
| | Z4 | 0.11 | 0.05 | 0.1 | 0.1 | 0.14 | 0.16 | 0.15 | 0.16 |
| | Z5 | 0.09 | 0.04 | 0.08 | 0.08 | 0.1 | 0.12 | 0.11 | 0.12 |
| | N1 | 0.04 | 0.03 | 0.06 | 0.05 | 0.03 | 0.04 | 0.03 | 0.04 |
| | N2 | 0.11 | 0.05 | 0.17 | 0.04 | 0.16 | 0.17 | 0.16 | 0.17 |
| | N3 | 0.2 | 0.04 | 0.25 | 0.05 | 0.26 | 0.24 | 0.25 | 0.24 |
| | N4 | 0.4 | 0.46 | 0.17 | 0.45 | 0.53 | 0.55 | 0.54 | 0.56 |

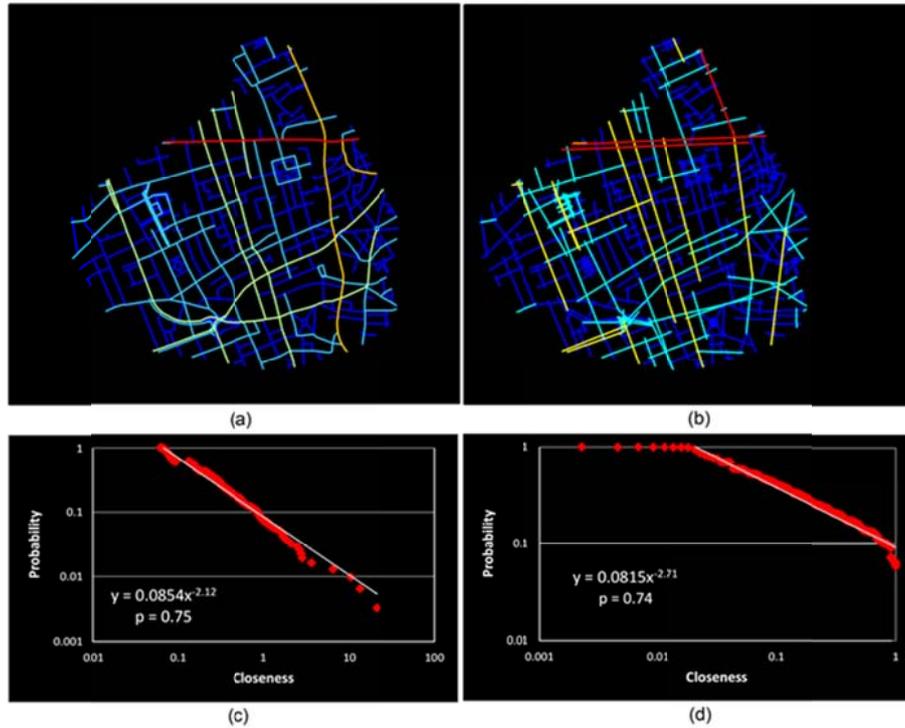

Figure 6: (Color online) The natural streets (a) and axial lines (b) in Z1 area and the power law distribution plot of the summed segment closeness values based on the point-based graph through natural streets (c) and axial lines (d)

To further support our argument that topology matters in predicting human activities, we conducted segment analysis and calculated other centrality measures (betweenness and closeness) under different kinds of segment-segment geometric relationships in terms of minimum length, least angle and fewest turns. All of the correlations between segment metrics and number of tweet locations are remarkably low (R square < 0.1). However, the correlations are significantly improved (R square > 0.5) when we summed of the centrality measures of segments into individual natural street. Thus far, this paper has offered both theoretical and empirical evidence to support the topological analysis for predicting human activities, and demonstrated how the scaling of far more the less-connected than the well-connected shapes human activities. We have shown that natural streets are the best representation at the city scale, followed by axial lines, and neither street nor line segments are able to capture human activities.

### 5. Further discussions on the implications of this study

In this section, we clarify the importance of space syntax or the topological analysis in general. As mentioned earlier, this paper is mainly motivated by the previous work by Hillier and Iida (2005) and further stimulated by the critique on space syntax (Ratti 2004). A common theme of those two papers is their geometric thinking, focusing on locations, distances, and directions. For a long time, space syntax has been criticized for its lack of geometric or metric information. As a matter of fact, it is the lack of geometric or metric information that gives space syntax its power to see the underlying scaling. The following discussions will surround the geometric representations and discuss why they are not as effective as topological representations for understanding human activities, or urban structure and dynamics in general.

Critics tend to say that space syntax discards precious metric information, so it is less useful for spatial analysis (Ratti 2004). To a great extent, the London underground map discards metric information, or its metric information has been dramatically distorted in order to show clearly station-station relationship. After or through the geometric distortion, the station-station topology is clearer than before. In this paper, we concentrate on street-street topology by discarding metric information, in order to see the underlying scaling of far more less-connected streets than well-connected ones. In other words, it is through discarding metric information or by concentrating on topological aspect of streets that we are able to capture the underlying scaling or living structure. As shown in this paper and many other studies in the space syntax literature, the topological analysis is capable of capturing a majority of human activities without considering metric information like street lengths and widths, and even building heights.

The traditional GIS analysis is largely based on the spatial models or representations that contain merely metric information and are therefore mechanistic. For example, a digital elevation model is a mechanical structure of numerous pixels, but the pixel-pixel-adjacent relationship says nothing about the underlying scaling structure of space. Geographic objects in the vector format can be identified from the imagery, but the topological relationship among the points, lines, and polygons still fails to uncover the underlying scaling hierarchy of far more small things than large ones. Therefore, as a critique to current geospatial models such as vector and raster, the geometric primitives of points, lines, polygons and pixels are mechanistically imposed, which means they pose little meanings in our perception and cognition. On the contrary, both natural streets and axial lines represent meaningful units in our cognition, and they are essentially small spaces that can be perceived form single viewpoints. Thus, the street-based topological analysis differs fundamentally from these GIS representations. In addition, GIS representations make no difference between space where people can move freely around and space occupied by buildings or other geographic features. The topological analysis concentrates only on spaces between buildings or, equivalently, streets in which people can freely move around. It is essentially the scaling hierarchy of the free space that shapes the human movement, or makes traffic predictable.

To this point, we can remark that human activities as a whole cannot be well understood from individual human beings. This is because the whole, as a complex system, is more than the sum of its parts. To borrow the famous statement by Winston Churchill – that *"we shape our buildings, and thereafter they shape us"* – we have demonstrated through the above case studies that human activities are substantially shaped by the underlying scaling of space. It is the underlying scaling structure that makes human activities predictable. However, this predictability is at a collective scale in terms of how many people come to individual streets, rather than detailed travel routes of individual people. Researchers tend to confuse the collective traffic flow shaped by the underlying street structure with individual travel behavior. This is also a common misunderstanding in the space syntax literature, as pointed out by Omer and Jiang (2015). In other words, the traffic flow captured by axial-based space syntax has little to do with human beings or random walkers. In a given street network, up to 80 percent of traffic flow can be accounted for by the underlying street structure (e.g., Jiang and Jia 2012). This social physical perspective (Buchanan 2007) is what underlies the topological representation and analysis.

**6. Conclusion**
Human activities are mainly shaped by the underlying street network or, more specifically, the scaling of numerous least-connected, a very few most-connected, and some in between the least- and most-connected. This is the major theme of the present paper, in which we have shown that natural streets are the best spatial representation, followed by axial lines, and that neither street segments nor line segments can well capture Tweet locations. Both natural streets and axial lines – or, more specifically, their topological representations – capture well the underling scaling hierarchy of far more the less-connected than the well-connected. On the other hand, street segments or line segments (more specifically, their relationships) do not reveal any scaling hierarchy or regularity. This is the major difference between geometric and topological representations. To reiterate, topology matters in predicating human activities, and in better understanding urban structure and dynamics.

The topological analysis relies on its unique topology of geometrically coherent units to reveal the underlying scaling of street structure. The scaling makes topological analysis of natural streets or axial lines-based space syntax unique and powerful. Human activities captured by the kind of topological analysis based on natural streets or axial lines are not individual travel behavior – which varies from person to person and from time to time for a single person – but collective travel behavior determined by the underlying scaling structure. In this paper, we have provided empirical evidence to show that topological representation with little metric information tends to capture a majority of human activities, whereas segment analysis with more metric information fails to capture human activities. This finding further supports our understanding that any spatial cognition derived from space syntax or the topological analysis is, at an aggregate level, determined by space rather than by people. Based on the understanding of the scaling or living structure of space, our future work will seek to develop new means for urban planning and design, e.g. how to generate or improve the scaling or living structure.

**Acknowledgement**
We initially developed this paper and had exactly the same conclusions as they are now while experimenting with traffic data of Dublin, Ireland. Unfortunately, the traffic data cannot be made public, so we changed it to the London Tweet data that are shared in ResearchGate (https://www.researchgate.net/publication/325812525_WhytopologyPaper). It is the reason that we used human activities with an implication on traffic flow. It is also the reason that Tweet users' activities reflect those of real people, which are largely shaped by the underlying scaling or living structure. An earlier of this paper was presented at ILUS 2017: International Land Use Symposium 2017, November 1–3, 2017, Dresden, Germany. Many thanks for anonymous reviewers for their constructive comments.